\documentclass{epl}
\title{Spin diffusion in doped semiconductors: the role of
Coulomb interactions}
\author{Irene D'Amico\inst{1} \and Giovanni Vignale\inst{2}}
\institute{
  \inst{1} Institute for Scientific Interchange (ISI) - Viale Settimio Severo 
65, 10133 Torino, Italy\\
  \inst{2} Department  of Physics,   University   of
Missouri, Columbia, Missouri 65211
}

\begin{document}

\maketitle

\begin{abstract}
We examine the effect of  Coulomb interaction on the mobility and diffusion
of  spin packets in  doped semiconductors. We find that  the diffusion
constant is {\it reduced}, relative to its non-interacting value, by the
combined effect  of Coulomb-enhanced spin susceptibility and spin Coulomb
drag.
In ferromagnetic semiconductors, the spin diffusion constant vanishes at the
ferromagnetic transition temperature.
\end{abstract}

The  ability to control inhomogeneous distributions of electrons
and holes in  semiconductors is essential to the operation of modern
electronic devices.  Unlike electron packets in  metals, which spread
out very quickly under the action of their own electric field, electron-hole
packets in semiconductors are  charge-neutral objects
and can therefore be long-lived.
The time evolution of such  packets
is controlled by a drift-diffusion equation with  mobility
and diffusion constants $\mu$ and $D$, as  was verified in
detail in the classic Haynes-Shockley experiment \cite{Shockley}.

Recently, a broader category of possible disturbances,  involving
inhomogeneous {\it spin} distributions, has come into sharp focus in the
context
of the emerging field of ``spintronics" \cite{Prinz}. Consider, for example, a
spin packet  consisting of excess  up-spin electrons compensated by
a deficiency of down-spin electrons. Such a disturbance can occur in the
conduction band of a metal or of a doped semiconductor \cite{Awscha1}.
Like an ordinary electron-hole packet this  is a charge-neutral
object  and can
therefore be extremely long-lived (recent experiments measure a spin
relaxation time $\tau_s$ of the order of $10~ns$ \cite{Awscha1,Awscha}).
Unlike an electron-hole
packet, however, the disturbance involves carriers of a single polarity
 - electrons  -  and therefore evolves with the mobility and
diffusion constants of the conduction band, which are usually larger than
those of the valence band. Indeed, large values of the spin drift mobility
$\mu_s
\sim 3 \times 10^3 cm^2/V s$ and spin diffusion constant  $D_s \sim 10^3
cm^2/s$ have been recently observed in experiments on n-doped GaAs
\cite{Awscha}.

The qualitative difference between unipolar and bipolar
disturbances in semiconductors  has  recently been emphasized by Flatt\'e and
Byers \cite{Flatte} within the frame of a simple model in which the
electron-electron interaction is treated in the Hartree approximation.
In this Letter, we refine their analysis  by examining
the more subtle effects of exchange and  Coulomb correlation on the
mobility and diffusion constant
of a spin packet.
It will be shown that these many-body effects have a serious impact
on the value of the spin diffusion constant and can be quantitatively
probed in a Haynes-Shockley-type experiment that measures independently
the mobility and the diffusion constant of a spin packet.  The two key physical
effects are (i) the reduction of the  spin stiffness (inverse of the spin
susceptibility) due to (mainly) exchange interactions
 and (ii)
the spin Coulomb drag \cite{spindrag}  working like friction against
the relative motion of up-spin and down-spin electrons.  Both effects
tend to reduce the diffusion constant. By contrast, the spin-packet
mobility turns out to be essentially unaffected by interactions.

Under certain conditions, the electron (hole) gas in doped semiconductors may
undergo a ferromagnetic transition.   The phenomenon occurs either at very
low densities, due to the Coulomb interaction \cite{Ceperley} or, for instance,
in $GaAs$ \cite{GaMn,Ohno} under heavy doping with magnetic $Mn$ impurities. As
the ferromagnetic transition temperature $T_c$ is approached   the
longitudinal spin stiffness vanishes and so does the spin diffusion constant,
which thus exhibits a critical behavior.

We begin our analysis by assuming, as usual, a   linear relationship
between the   number current densities  $\vec J_\alpha (\vec r)$ ($\alpha =
\uparrow$ or $\downarrow$) and the gradient of the local electro-chemical
potentials
$\psi_\alpha (\vec r) = \phi(\vec r) - (1 / e) \partial f (n_{\uparrow},
n_{\downarrow},T) / \partial n_\alpha$,
where $\phi(\vec r)$ is the electrostatic potential, $e$ is the absolute
value of the   electron charge, and   $f (n_{\uparrow}, n_{\downarrow},T)$
is the free energy per unit volume of a homogeneous interacting electron gas at
the local spin densities $n_{\alpha} (\vec r)$ and temperature $T$
\cite{Footnote0}.    This leads to the  equation
 \begin{equation}
\label{driftdiffusion} e  \vec J_\alpha (\vec r)  =   \sum_{\beta} \left (
\sigma_{\alpha \beta}  (\vec r) \vec \nabla \phi  (\vec r) - e D_{\alpha
\beta} (\vec r) \vec \nabla n_{\beta} (\vec r) \right )
\end {equation}
where   $\sigma_{\alpha \beta}$ is  the homogeneous conductivity matrix,
calculable from the Kubo formula
and the diffusion matrix $D_{\alpha \beta}$ is given by the generalized
Einstein relation
\begin{equation}
\label{diffusionmatrix}
e^2  D_{\alpha  \beta} = \sum_\gamma \sigma_{\alpha \gamma}
S_{\gamma \beta}
\end{equation}
where
\begin{equation}
\label{spinstiffness}
S_{\alpha \beta} =
{\partial^2 f (n_{\uparrow}, n_{\downarrow},T) \over \partial n_{\alpha}
\partial n_{\beta}}
\end{equation}
is the static spin-stiffness matrix  - minus the inverse of the spin
susceptibility  matrix.

On a formal level  the main effect of the Coulomb interaction is
the appearance of non-vanishing off-diagonal elements of the conductivity and
spin-stiffness matrices.  These off-diagonal matrix elements have a simple
physical interpretation.  $\sigma_{\uparrow \downarrow} \neq 0$ implies that an
electric field acting only on the up-spin electrons must necessarily drag along
a current of down-spin electrons.  Similarly, $S_{\uparrow \downarrow} \neq 0$
means that the chemical potential of up-spins $\partial f  (n_{\uparrow},
n_{\downarrow},T)  /\partial n_{\uparrow}$ is a function
of {\it both} up and down spin densities.  In addition the Coulomb
interaction significantly modifies the values of the diagonal elements of these
matrices, as we shall see momentarily.

Let us apply eq.~(\ref{driftdiffusion}) to the calculation of the time
evolution of
a spin packet obtained by injecting an {\it excess} spin density $\Delta
m(\vec r, 0) = \Delta M \delta (\vec r)$ near the origin at time $t=0$. We
denote by $ m (\vec r,t) = n_{\uparrow} (\vec r,t) - n_{\downarrow} (\vec
r,t)$ the net spin density at point $\vec r$ and time $t$, by $m^{(0)} =
n_{\uparrow}^{(0)} - n_{\downarrow}^{(0)}$ the uniform value of the spin
density at thermodynamic equilibrium, and by $\Delta m( \vec r ,t) \equiv
m(\vec r,t) -m^{(0)}$ the excess spin density following spin injection.
Similarly we denote by $ n (\vec r,t) = n_{\uparrow} (\vec r,t) +
n_{\downarrow} (\vec r,t)$  the total electron density and by $n^{(0)}$
and by $\Delta n (\vec r,t)$ the equilibrium and the excess density
respectively.

The solution of the problem requires as additional inputs
the
continuity equations for the number and spin densities
 \begin{eqnarray}
\label{chargecontinuity}
{\partial \Delta n (\vec r,t) \over \partial t} &=& - \vec \nabla \cdot \vec
J(\vec r,t)\\
 \label{spincontinuity} {\partial \Delta m(\vec r,t) \over
\partial t} &=& - {\Delta m (\vec r,t) \over \tau_s} - \vec \nabla \cdot
\vec J_m(\vec r,t), \end{eqnarray} where $\vec J_m = \vec J_{\uparrow} -
\vec J_{\downarrow}$ is the  spin  current density and $\tau_{s}$
is the spin relaxation time, which is very long \cite{Awscha1,Awscha}.

In practice, following the procedure familiar in the theory of bipolar
carrier packets \cite{Smith}, we first combine eqs.
(\ref{chargecontinuity}),
(\ref{spincontinuity}) and (\ref{driftdiffusion}) to eliminate the  $\vec
\nabla \cdot \vec E$ term
 related to the Poisson equation,
and  {\it then} impose the local charge
neutrality constraint \cite{Footnote1}
\begin{equation} \label{chargeneutrality}
\Delta n_{\uparrow}(\vec r,t) = - \Delta n_{\downarrow}(\vec r,t).
\end{equation}
This yields the result
\begin{eqnarray}
\label{packeteom}
{\partial \Delta m(\vec r,t) \over \partial t}& = & - {\Delta m (\vec r,t)
\over
\tau_s} + D_s \nabla^2  \Delta m(\vec r,t)\nonumber
\\ & ~ & + \mu_s \vec E \cdot \vec \nabla
\Delta m(\vec r,t)
\end{eqnarray}
where
\begin{equation}
\label{muspin}
 \mu_s =  {(n_{\uparrow}+n_{\downarrow})\tilde \sigma_{\uparrow} \tilde
\sigma_{\downarrow} \over n_{\uparrow}n_{\downarrow} ( \tilde
\sigma_{\uparrow}+
\tilde \sigma_{\downarrow})}
\end{equation}
and
\begin{equation}
\label{Dspin}
 D_s =  {  \tilde \sigma_{\uparrow} \tilde D_{\downarrow} +
 \tilde \sigma_{\downarrow} \tilde D_{\uparrow} \over
  \tilde \sigma_{\uparrow} +   \tilde
\sigma_{\downarrow}}
\end{equation}
are the effective mobility and diffusion constants  \cite{Footnote2},
and $\vec E$
is an externally applied electric field.  Eqs.~(\ref{muspin}) and
(\ref{Dspin})
reduce to the expressions presented in \cite{Flatte} in the noninteracting
case.
The constants $\tilde \sigma_\alpha$ and $\tilde
D_\alpha$ are presently given by \begin{eqnarray}
\label{tildes}
\tilde \sigma_\alpha &=& \sigma_{\alpha  \alpha}+\sigma_{\alpha  \bar \alpha}
\nonumber \\
\tilde D_\alpha &=& D_{\alpha  \alpha}- D_{\alpha  \bar \alpha}.
\end{eqnarray}

The fact that the conductivities enter a spin symmetric combination while the
diffusion constants are in a spin antisymmetric combination reflects the fact
that the electrostatic field has the same sign for both spin components, while
the density gradients have opposite signs (see (\ref{chargeneutrality})).

The solution of eq.~(\ref{packeteom}) in a homogeneous and isotropic liquid
is
\begin{equation}
\label{packetevolution}
\Delta m  (\vec r,t) = {\Delta M e^{-t/\tau_s} \over (4 \pi D_s t)^{3/2}}
e^{-{|\vec r + \mu_s \vec E t|^2 \over 4 D_s t}}.
\end{equation}

Thus, we see that a Haynes-Shockley-type experiment can in principle determine
 $\mu_s$ and $D_s$ independently, provided that $\tau_s$ is sufficiently long.

Let us now proceed to the calculation of $\mu_s$ and $D_s$.
The necessary inputs are the conductivity and the spin stiffness matrices.
The
conductivity matrix is best computed from its inverse, namely the resistivity
matrix $\rho_{\alpha \beta}$ whose general structure is determined by the
principle of Galilean invariance and Newton's third law.  The explicit form of
$\rho_{\alpha \beta}$, extracted from  eq. (3) of Ref.(\cite{spindrag}), is
 \begin{equation}
\label{resistivity} {\bf \rho} = \left( \matrix{{m^* \over n_{\uparrow} e^2
\tau_{\uparrow}}-{n_{\downarrow} \over n_{\uparrow}} \rho_{\uparrow
\downarrow}&\rho_{\uparrow \downarrow}\cr \rho_{\uparrow \downarrow}& {m^*
\over
n_{\downarrow} e^2 \tau_{\downarrow}}-{n_{\uparrow} \over n_{\downarrow}}
\rho_{\uparrow \downarrow}} \right ).
\end{equation}
Here $\tau_{\alpha}$ are the combined electron-impurity and
electron-phonon scattering times,  usually of the order of $10^{-3} -
10^{-4}~ns$,  $m^*$ is the effective mass of the carriers, and  $\rho_{\uparrow
\downarrow}$ is the spin drag transresistivity calculated in
Ref.~(\cite{spindrag}).  After computing  $\sigma_{\alpha \beta} =
[\rho^{-1}]_{\alpha \beta}$ the diffusion matrix is
straightforwardly   obtained from eq. ~(\ref{diffusionmatrix}), and $\mu_s$,
$D_s$ are calculated from eqs.~(\ref{muspin}) and (\ref{Dspin}).
The algebra  is greatly simplified by the
reasonable assumption that the scattering times  for the two spin components
are not too different, i.e.,   $\tau_{\uparrow} = \tau_{\downarrow} =
\tau_D$ \cite{Footnote3}.
Under this condition, the result is
\begin{equation}
\label{musfinal}
\mu_s = {e \tau_D  \over m^*}
\end{equation}
and
\begin{equation}
\label{dsfinal}
D_s = {\mu_s k_BT \over e}  {S \over S_c}  {1 \over 1 -
\rho_{\uparrow \downarrow}/\rho_D},
\end{equation}
where $S =  \partial^2 f(n,m,T)/\partial m^2$ is the spin stiffness, $S_c =
k_BTn/4n_{\uparrow}n_{\downarrow}$ is the Curie spin stiffness of an ideal
classical gas, and $\rho_D =m^*/ ne^2 \tau_D$ is the  Drude
resistivity.

Eq.~(\ref{musfinal}) tells us that the mobility of the packet is not
{\it explicitly} modified by the Coulomb interaction and in fact coincides with
the ordinary homogeneous mobility.  Strictly speaking this result is only
valid under the assumption that  up-spin and down-spin electrons have equal
mobilities and thus drift at the same speed in an applied electric field.
Coulomb interactions, being Galilean-invariant,  cannot change the total
momentum of such a uniformly drifting electron gas.

The situation is completely different for the diffusion constant. As the  spin
packet spreads out the up- and down-spin currents are  directed in opposite
directions and friction arises: for this reason the expression for $D_s$
contains the spin-drag resistivity as a factor that reduces the numerical value
of $D_s$.  In addition,  the Coulomb interaction together with the Pauli
exclusion principle  reduces the energy cost of spin-density fluctuations
(i.e., the spin stiffness)  further decreasing the rate of
diffusion of a spin packet.

Fig.1 presents the necessary ingredients to calculate $D_s$.
Fig.1a shows that   $\rho_{\uparrow \downarrow}$ - a negative number - vanishes
at low temperature as $(T/T_F)^2$ 
peaking at a temperature of the order of the Fermi temperature $T_F$. 
As the inset illustrates, the prefactor $1/(1-
\rho_{\uparrow \downarrow}/\rho_D)$ (see eq.~(\ref{dsfinal})) 
displays a marked dependence on the sample
mobility,
 increasing with the latter.
Fig.1b shows $S$ rescaled
by its non-interacting value $S_{ni}$ and its behavior at the onset of the
 ferromagnetic instability. We evaluated $S$ numerically starting  from   the
work of Tanaka and Ichimaru \cite{Ichimaru} where the  free energy density
is calculated as a function of temperature, density, and spin polarization,  by
means of a self-consistent  integral equation approach  that
satisfies the thermodynamic sum rules.

In fig.2 we plot   $D_s/D_{c}$, where $D_c = \mu_s
k_BT/e$ is the classical noninteracting diffusion constant,  for n-doped
GaAs in
a range of densities that are relevant to  the experiments of Ref.
\cite{Awscha}. The solid line corresponds to our full-interacting calculation,
while the dashed line to the non-interacting case.
We see that the interaction correction is quite significant, and reduces the
value of $D_s$ as expected. Despite this reduction, $D_s$ is still
considerably larger than $D_c$  consistent with  experimental observations.
These results show that the effect of the reduced spin stiffness dominates at
low and intermediate temperature, while the spin drag contribution dominates
 at high temperature (see inset of fig.~\ref{fig2}).

In the non-degenerate limit $T>>T_F(n)$ we find that $D_s/D_{c}$ approaches
$1$.
In the noninteracting theory \cite{Flatte} this limit is approached from above
because the leading correction to the spin stiffness coming from the quantum
kinetic energy  is positive: $S_{ni} = S_c[1 +\lambda^3_Tn/2\sqrt{2}]$ where
$\lambda_T$ is
the De Broglie thermal wavelength at temperature $T$.  In the interacting
theory
instead,
the leading correction to $D_s$ is due to the spin Coulomb drag term and it
is negative.
In fact, for $T>>T_F(n)$,
$\rho_{\uparrow \downarrow}/ \rho_D \sim [n/ (k_BT)^{3/2}]
\ln (n/ (k_BT)^{2})$,
and the logarithmic term dominates over corrections entering the spin
stiffness in both the $n \to 0$ and $T\to\infty$ limits.
Due to  interactions, $D_s /D_c \to 1$ from {\it below} always.

A very  interesting feature of eq.~(\ref{dsfinal}) is the
possibility of
large variations in $D_s$ when the electron gas undergoes a ferromagnetic
transition.  From the curves in fig.1b and fig.3  we see that $S$
(interpreted
as {\it longitudinal} spin stiffness in the ferromagnetic phase)  and $D_s$
vanish at the transition temperature $T_c$. For $T<T_c$, $D_s/D_c$ increases at
first, due to the sharp increase in spin stiffness (see fig.1), but then begins
to saturate and tends to 1 as full polarization sets in \cite{footnote4}.

In an ordinary electron liquid, the ferromagnetic transition is
predicted  to occur only at extremely low densities \cite{Ceperley}.  There is,
however, an interesting variant:  semiconductors doped with magnetic
impurities (Mn)  can undergo a ferromagnetic transition at rather high carrier
densities \cite{Ohno,Konig}, $n \sim 10^{20} cm^{-3}$ for (Ga,Mn)As, and
temperatures as large as $110 K$ \cite{Ohno}.  Our theory can be extended to
such materials, but, for the sake of clarity, we shall present this
extension in a longer paper.  Here we simply remark that in a simple mean field
theory, such as that of Ref. (\cite{Jungwirth}), the calculated value of
$D_s/D_c$ for realistic values of the parameters has the form shown in fig. 3b,
where,   as in the intrinsic ferromagnetic case, $D_s$ vanishes at $T_c$
\cite{footnote5}.

In conclusion, we have demonstrated that a Haynes-Shockley experiment
measuring   $D_s$ and $\mu_s$ for a unipolar spin packet would be a sensitive
probe of many-body effects such as the spin-Coulomb drag and the Coulomb
enhancement of the spin susceptibility, and would provide a strong signature
of a ferromagnetic ordering transition.  Conversely, many-body effects must be
taken into account in a quantitative theory of spin diffusion.

\acknowledgments
We gratefully acknowledge support from NSF grants  DMR-9706788 and DMR-0074959.

\begin{figure}
\onefigure{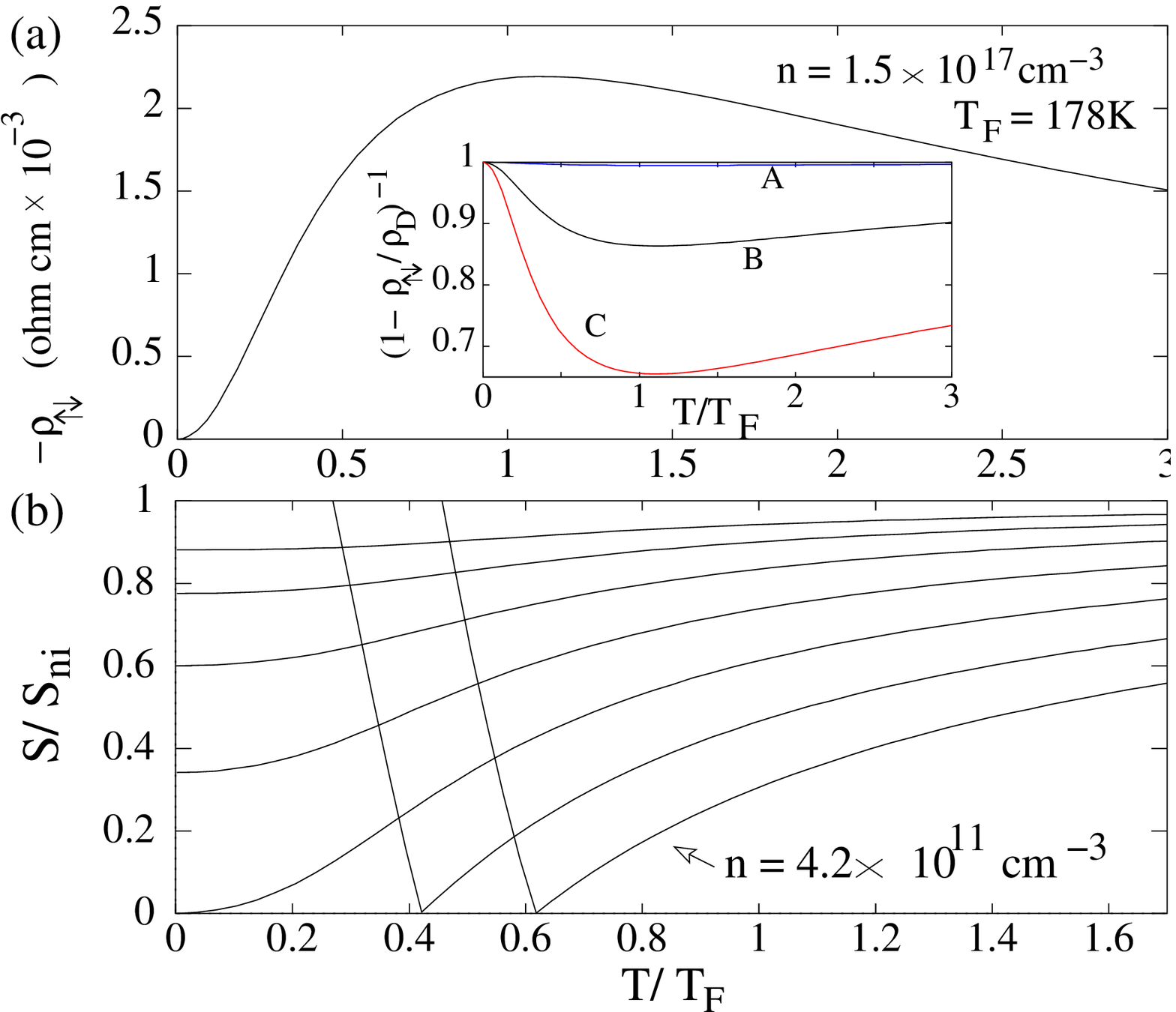}
\caption{
(a) Transresistivity $\rho_{\uparrow
\downarrow}$ vs the reduced temperature $T/T_F$ for typical semiconductor
parameters.  The inset shows the behavior of the factor $1/(1-
\rho_{\uparrow \downarrow}/\rho_D)$ for three different mobilities: $\mu=10^2
cm^2/Vs$ (A), $\mu=3\times 10^3
cm^2/Vs$ (B),  $\mu=10^4
cm^2/Vs$ (C).
 (b) Spin stiffness $S$ vs $T/T_F$. The  density is $n=4.2\times 10^{11}cm^{-3}$ for the lower
curve and increases by a factor $10$ for each line starting from the bottom.
The cusps
represent the onset of   ferromagnetism.
}
\label{fig1}
\end{figure}

\begin{figure}
\onefigure{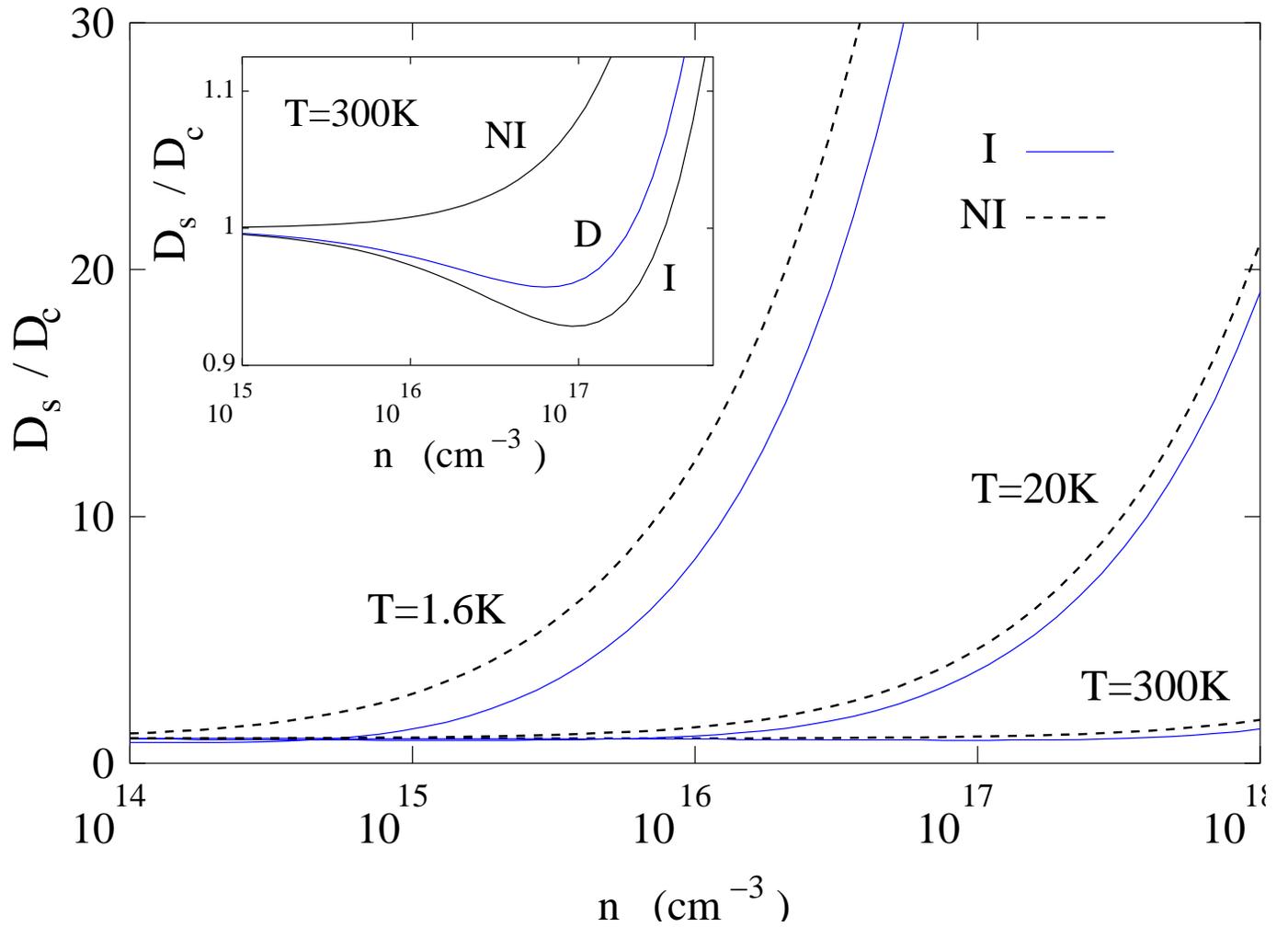}
\caption{
The interacting diffusion constant of a spin-packet (I) and its
non-interacting
approximation (NI) (rescaled by  $D_c$) vs
 density for different temperatures.
The inset shows the comparison with the value obtained considering
 interactions only through the spin Coulomb drag effect (D). In all the
calculations
the dielectric constant of the semiconductor is $\epsilon =12$ and the
mobility is $\mu = 3\times 10^3 cm^2/Vs$.
}
\label{fig2}
\end{figure}

\begin{figure}
\onefigure{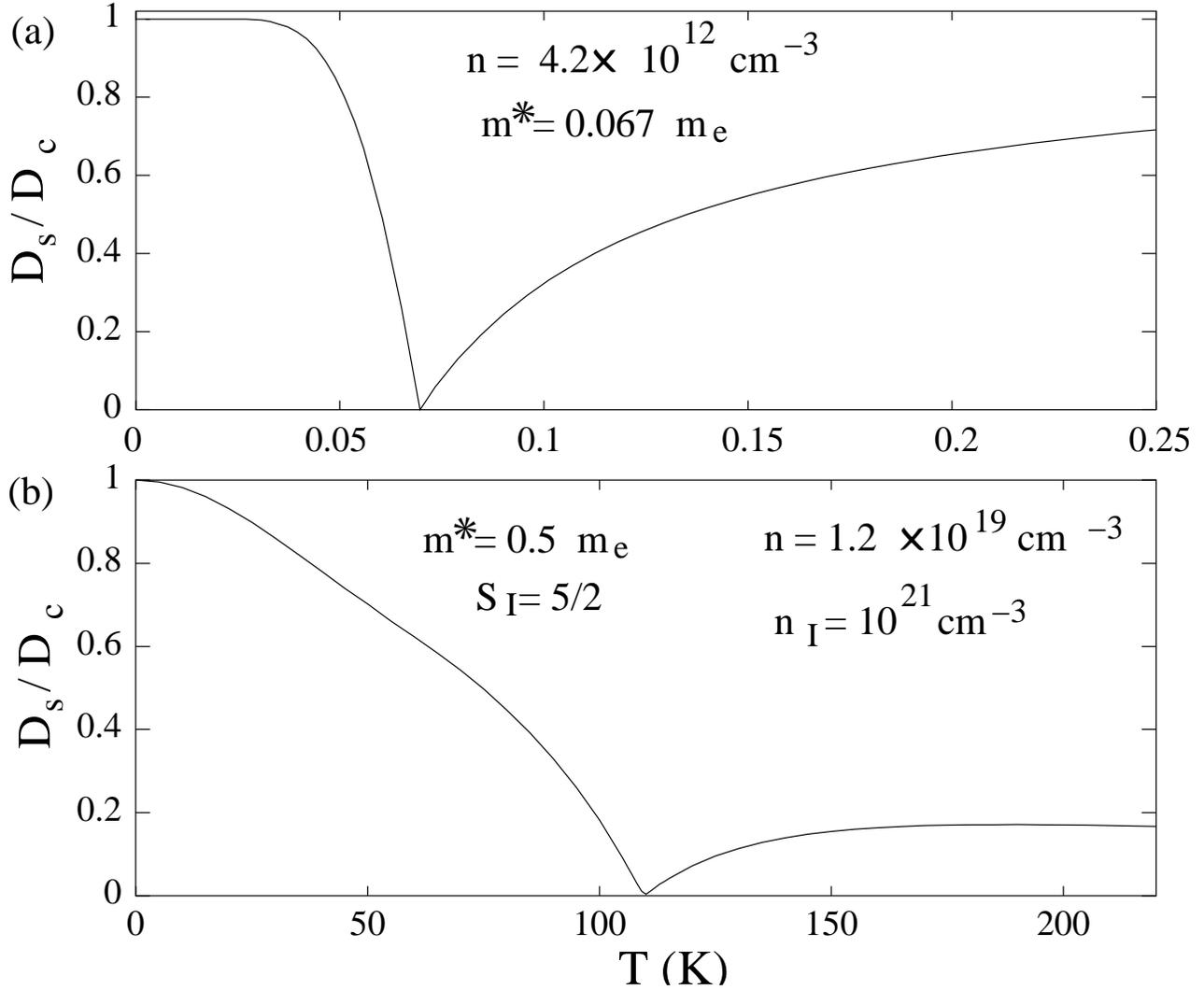}
\caption{
(a) The diffusion constant of a spin packet vs temperature in a low-density
electron gas  with no magnetic impurities. (b) Same as (a) for an electron
gas of typical density $n=1.2 \times 10^{19} cm^{-3}$ in a semiconductor
doped with magnetic impurities ($n_I$ and $S_I$ are the impurity
concentration and spin respectively).
}
\label{fig3}
\end{figure}




\begin{thebibliography}{0}
\bibitem{Footnote0}  We assume that impurities have a negligible effect on the
thermodynamic functions of the electron gas.
\bibitem{Footnote1} Any excess charge is spread out to infinity within a
dielectric relaxation time ($\sim 10^{-4} ns$), which is much shorter
than the time scale of interest here.
\bibitem{Footnote2}  In the spirit of the linear response approximation these
constants should be evaluated at the underlying equilibrium density of the
electron gas.
\bibitem{Footnote3} This assumption is well justified for non-degenerate
carriers.  In the degenerate case $1/\tau_\alpha$ has a weak $n_\alpha^{1/3}$
dependence on the density.
\bibitem{footnote4} If spin polarization is full for $T \to 0$ the spin
diffusion constant reduces to the diffusion constant of electrons of minority
orientation which has the classical value $D_c = k_BT \mu/e$   in this regime.
\bibitem{footnote5}   This result holds under the assumption that
the magnetic impurities are able to adjust their magnetization to the local
excess spin density of the electron gas within the diffusion time scale.
Otherwise,  the diffusion constant does not vanish, but has a kink at $T_c$.
\end{thebibliography}
\end{document}